\documentstyle[psfig]{mn}
\title{A study of the Galactic plane towards $l = 305^{\circ}$}
\author[Baume  et al.]
{G. Baume$^1$,\thanks{Member of Carrera del Investigador CONICET, Argentina}
 G. Carraro$^2$\thanks{On leave from Dipartimento di Astronomia, Universit\`a di Padova, Vicolo Osservatorio 2, I-35122, Padova, Italy} and
 Y. Momany$^2$\thanks{On leave from INAF, Osservatorio Astronomico di Padova, Vicolo Osservatorio 2, I-35122, Padova, Italy}
  \thanks{email: gbaume@fcaglp.fcaglp.unlp.edu.ar (GB),
                gcarraro@eso.org (GC)
                ymomany@eso.org (YM)} \\
      $^1$Facultad de Ciencias Astron\'omicas y Geof\'{\i}sicas (UNLP),
          Instituto de Astrof\'{\i}sica de La Plata (CONICET, UNLP), \\
          \hspace{0.2cm}Paseo del Bosque s/n, La Plata, Argentina \\
      $^2$ESO, Alonso de Cordova 3107, Vitacura, Santiago de Chile, Chile \\
}
\date{\it Submitted: *** 2009}
\pubyear{2009}
\begin{document}
\maketitle
\title{Galactic plane at $l = 305^{\circ}$}

%%%%%%%%%%%%%%%%%%%%%%%%%%%%%%%%%%%%%%%%%%%%%%%%%%%%%%%%%

\begin{abstract} We present optical  ($UBVI_C$) observations of a rich
  and complex field in the Galactic plane towards $l \sim 305^{\circ}$
  and $b  \sim 0^{\circ}$. Our  analysis reveals a  significantly high
  interstellar absorbtion  ($A_V \sim 10$) and  an abnormal extinction
  law in this line of  sight.  Availing a considerable number of color
  combinations,  the  photometric  diagrams  allow us  to  derive  new
  estimates  of the fundamental  parameters of  the two  open clusters
  Danks~1 and  Danks~2. Due to  the derived abnormal reddening  law in
  this line  of sight, both clusters  appear much closer  (to the Sun)
  than previously thought.
  Additionally, we  present the optical  colors and magnitudes  of the
  WR~48a star and its main parameters were estimated. The properties
  of the two embedded clusters DBS2003~130 and 131, are  also addressed.
  We identify a  number of Young Stellar Objects which are probable
  members of  these  clusters. This new material is then used to revisit
  the spiral structure in this sector of the  Galaxy showing evidence
  of populations associated with the inner Galaxy Scutum-Crux arm.

\end{abstract}


\section{Introduction}

The study of embedded Galactic  clusters is fundamental in tracing the
Galactic spiral  structure and improve  our understanding of  the star
formation process.  Our group  has investigated several Galactic plane
regions in  the fourth quadrant line  of sights (see  V\'azquez et al.
2005, Carraro \& Costa 2009), and in this paper we further extend this
study to a region located at $l = 305^{\circ}$ and $b \sim 0^{\circ}$.
This  region contains  very  interesting objects;  being an  important
cloud  which obscure  two  compact young  open  clusters (Danks~1  and
Danks~2), a  WR star  (WR~48a), and at  least three  embedded clusters
(DBS2003~130, 131 and 132) so far detected only in the infrared (Dutra
et al.  2003). There are  also several HII regions and OH/H$_2$O maser
sources (see Danks et al.  1984 or Clark \& Porter 2004 for a detailed
description).

The open clusters Danks~1 and  Danks~2 were first detected by Danks et
al. (1983) and  their parameters were only recently  estimated by Bica
et al. (2004), using $BVI$  and $JHK$ photometry.  In this regards, we
show  that  complementing  the  optical  and infrared  data  with  $U$
observations   adds   valuable  information   that   allow  a   better
determination of the interstellar absorption in this line of sight.
As  for the  embedded  clusters, it  is  important to  note that  only
DBS2003~131 has  been extnsively studied  in the infrared  (Leistra et
al. 2005 and  Longmore et al.  2007), and that none  of 3 clusters has
been studied in the optical.

In the  present paper we perform a  detailed wide field study  at $l =
305^{\circ}$  using all $UBVI$  and $JHK$,  of all  the aforementioned
objects,   and   derive  updated   estimates   of  their   fundamental
parameters. We then analyze this field with respect to other fields in
the fourth Galactic quadrant  to derive information about the Galactic
spiral structure of this portion of the disk.

The  layout of  the paper  is as  follows. In  Sect.~\ref{sec:data} we
describe   the  used   data,   and  the   reduction  and   calibration
procedures. In Sect.~\ref{sec:analysis} we present the analysis of the
data  together   with  the  main  clusters   parameters.  Finally,  in
Sects.~\ref{sec:discussion}  and \ref{sec:conclusions} we  discuss and
summarize our results.

\begin{figure*}
\begin{center}
\centerline{\psfig{file=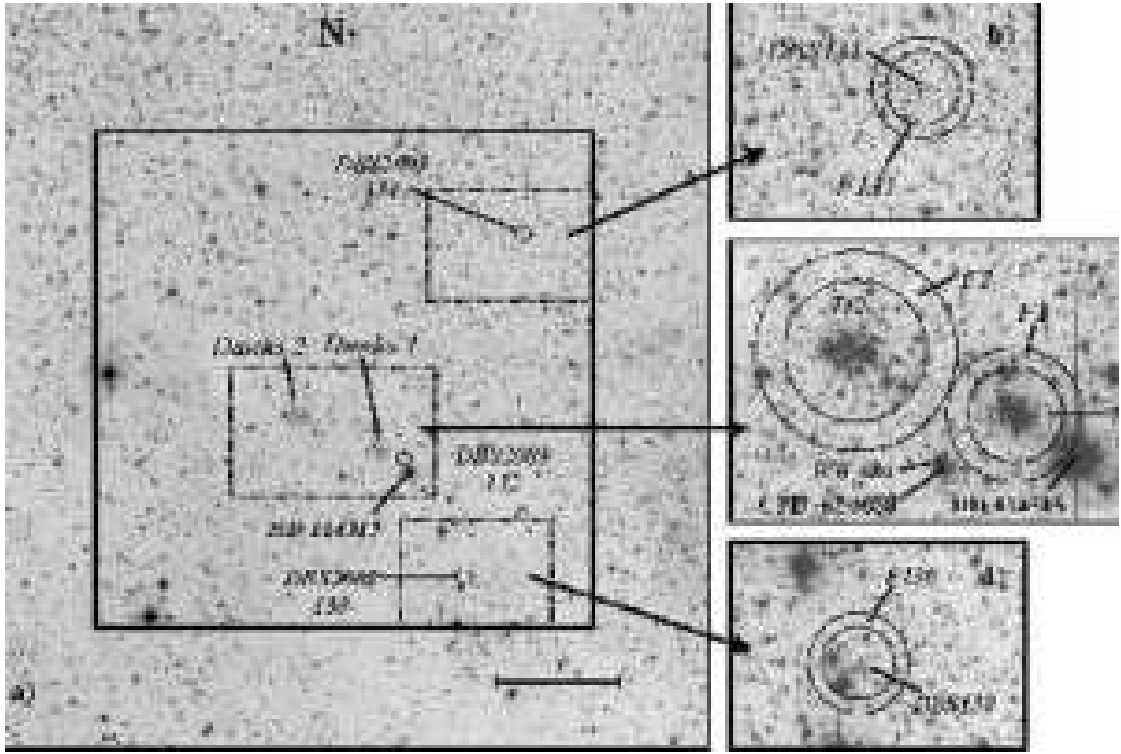,width=15cm}}
\caption{Left image (panel a)  shows a second generation Digitized Sky
  Survey  (DSS-2;  red  filter)  image centered  at  $\alpha_{2000}  =
  13:12:43.4$; $\delta_{2000}  = -62:39:2.2$  and covering a  field of
  $30\farcm0  \times 30\farcm0$.   Open  clusters Danks~1/2;  embedded
  clusters  DBS2003~130,   131  and  132;  and   HD~1145115  star  are
  identified.  Solid  square indicates  the full studied  area whereas
  the  dotted rectangles trace  the regions  highlighted in  the right
  minor  panels  (b,  c and  d),  as  these  appear  on our  long  $I$
  exposure.}
\label{fig:dss}
\end{center}
\end{figure*}

\section{Data} \label{sec:data}

\subsection{Observations}

$UBVI_C$  images of  the region  under study  (see Fig.~\ref{fig:dss})
were acquired using the Y4KCAM  camera attached to the 1.0m telescope,
which      is     operated      by      the     SMARTS      consortium
\footnote{http://www.astro.yale.edu/smarts} and placed at Cerro Tololo
Inter-American Observatory (CTIO). This camera is equipped with an STA
$4064  \times 4064$  CCD with  $15 \mu$  pixels. This  set-up provides
direct imaging over a field of view (FOV) $20\farcm0 \times 20\farcm0$
with a scale of $0\farcs289$/pix.  The relatively large FOV allowed us
to include several  objects present in the region and  also to have an
important sample  of the  adjacent stellar field  in this part  of the
Galactic plane.   The CCD was  operated without binning, at  a nominal
gain and  read out noise  of 1.44 e-/ADU  and 7 e- per  quadrant (this
detector  is  read  by  means  of four  different  amplifiers).  Other
detector       characteristics       can       be       found       at
http://www.astronomy.ohio-state.edu/Y4KCam/detector.html.   Details on
the observations are given  in Table~\ref{tab:frames}. Typical FWHM of
the  data  was  about  $0\farcs9$  and  airmasses  values  during  the
observation of the scientific frames ranged 1.19-1.22.

\begin{table}
\fontsize{8} {10pt}\selectfont
%\tabcolsep 0.20truecm
\caption{Journal of observations of the scientific frames together with
used calibration coefficients (23-24 March 2006).}
\begin{tabular}{clccccc}
\hline
\multicolumn{7}{l}{Exposure times [sec]x$N$} \\
\hline
Date & \multicolumn{1}{l}{Frames} & $U$ & $B$ & $V$ & $I_C$ & $N_f$ \\
\hline
\multicolumn{6}{l}{Scientific frames} \\
\hline
23-24 & long   & 2000x5 & 1200x2 & ~900x2 & ~700x2 & 1 \\
      & medium & ~200x2 & ~100x2 & ~100x2 & ~100x2 & 1 \\
      & short  & ~~30x2 & ~~30x2 & ~~30x2 & ~~30x2 & 1 \\
\hline
\multicolumn{7}{l}{Standard frames (Landolt 1992)} \\
\hline
23 & SA101 & ~400 & ~200 & ~150 & ~130 & 2 \\
   & SA107 & ~200 & ~~50 & ~~30 & ~~30 & 3 \\
24 & SA101 & ~400 & ~200 & ~150 & ~130 & 2 \\
   & SA101 & ~200 & ~~50 & ~~30 & ~~30 & 1 \\
   & SA104 & ~400 & ~200 & ~150 & ~130 & 1 \\
   & SA104 & ~200 & ~~50 & ~~30 & ~~30 & 1 \\
   & SA107 & ~200 & ~~50 & ~~30 & ~~30 & 3 \\
\hline
\hline
\multicolumn{7}{l}{Calibration and extinction coefficients} \\
\hline
\multicolumn {3}{l}{$u_1 = +3.284 \pm 0.007$} & \multicolumn {4}{l}{$v_{1bv} = +1.822 \pm 0.010$} \\
\multicolumn {3}{l}{$u_2 = -0.025 \pm 0.012$} & \multicolumn {4}{l}{$v_{2bv} = -0.054 \pm 0.012$} \\
\multicolumn {3}{l}{$u_3 = +0.45$}            & \multicolumn {4}{l}{$v_{1vi} = +1.836 \pm 0.008$} \\
\multicolumn {3}{l}{$b_1 = +2.025 \pm 0.012$} & \multicolumn {4}{l}{$v_{2vi} = -0.059 \pm 0.009$} \\
\multicolumn {3}{l}{$b_2 = +0.167 \pm 0.015$} & \multicolumn {4}{l}{$v_3= +0.16$}                 \\
\multicolumn {3}{l}{$b_3 = +0.25$}            & \multicolumn {4}{l}{$i_1 = +2.677 \pm 0.016$}     \\
\multicolumn {3}{l}{}                         & \multicolumn {4}{l}{$i_2 = -0.002 \pm 0.018$}     \\
\multicolumn {3}{l}{}                         & \multicolumn {4}{l}{$i_3 = +0.08$}                \\
\hline
\hline
\label{tab:frames}
\end{tabular}
\begin{minipage}{8.5cm}
\vspace{0.1cm}
{\bf Notes:} $N$ indicates the number of obtained exposures
in case it is more than 1, while $N_f$ indicates the amount of different
observed frames in each field.
\end{minipage}
\end{table}

\subsection{Reduction}

All   frames  were  pre-processed   in  a   standard  way   using  the
IRAF\footnote{IRAF is  distributed by NOAO, which is  operated by AURA
  under cooperative agreement with  the NSF.}  package CCDRED. To this
aim, zero exposures and sky flats were taken every night.  In order to
achieve deep photometry, all the  long exposures within each band were
combined using  IMCOMBINE task. This procedure helps  to remove cosmic
rays and improve  the signal-to-noise ratio of the  faintest stars. In
particular,  five images  were combined  in $U$  band, allowing  us to
reach  the  $U-B$  index  for  the  brightest  stars  of  Danks~1  and
Danks~2. However it was not  enough for the case of embedded clusters,
were only data in $BVI$ bands could be obtained.

Photometry  was   then  performed  using  IRAF   DAOPHOT  and  PHOTCAL
packages. Instrumental magnitudes were obtained using the point spread
function  (PSF) method  (Stetson 1987).   Since  the FOV  is large,  a
quadratic spatially  variable PSF was adopted and its calibration
on each image was done using several isolated, spatially well distributed,
bright stars (about 25) across the field. The  PSF photometry was
finally aperture-corrected for each filter and exposure time. Aperture
corrections were computed performing aperture photometry of a suitable
number (about 20)  of bright stars in the field. In  order to obtain a
more  complete  sample  of  the  stars  in  the  observed  region,  an
additional  photometry   table  was  generated  using   as  input  the
coordinates (converted to  pixels) of all the stars  that according to
the 2MASS  catalogue must  be present in  our FOV.  Finally,  all data
from  different filters  and  exposures were  combined and  calibrated
using DAOMASTER (Stetson 1992).

\subsection{Photometric calibration}

Standard  star selected  areas (see  Table~\ref{tab:frames})  from the
catalog of  Landolt (1992) were  used to determine  the transformation
equations  relating  our   instrumental  magnitudes  to  the  standard
$UBVI_C$ system.   The selection  of the fields  was done in  order to
provide a wide range in  colors. Then, aperture photometry was carried
out for  all the standard stars  ($\sim 70$ per night)  using the IRAF
PHOTCAL package.  To  tie our observations to the  standard system, we
use transformation
equations of the form:\\

\begin{center}
\begin{tabular}{llc}
$u = U + u_1 + u_2 (U-B) + u_3 X$           & ($r.m.s. = 0.04$) & (1) \\
$b = B + b_1 + b_2 (B-V) + b_3 X$           & ($r.m.s. = 0.03$) & (2) \\
$v = V + v_{1bv} + v_{2bv} (B-V) + v_3 X$   & ($r.m.s. = 0.02$) & (3) \\
$v = V + v_{1vi} + v_{2vi} (V-I_C) + v_3 X$ & ($r.m.s. = 0.02$) & (4) \\
$i = I_C + i_1 + i_2 (V-I_C) + i_3 X$       & ($r.m.s. = 0.02$) & (5) \\
\end{tabular}
\end{center}

\noindent  where $UBVI_C$  and  $ubvi$ are  standard and  instrumental
magnitudes respectively and $X$ is the airmass of the observation. The
transformation coefficients  and extinction coefficients  for the CTIO
Observatory  are  shown  at  the  bottom of  Table~1.  To  derive  $V$
magnitudes,  we  use  expression   (3)  when  the  $B$  magnitude  was
available; otherwise expression (4) was used.

\subsection{Complementary data and astrometry}

Other  available catalogues,  such as  the Two-Micron  All  Sky Survey
(2MASS; Cutri et al. 2003,  Skrutskie et al.  2006) are of fundamental
importance to  perform a  more complete analysis  of the  region under
investigation.  Therefore,  using the X--Y  stellar positions obtained
from our  data, their equatorial  coordinates were computed.  First of
all,  a  matched list  of  X--Y  and RA,  DEC  was  built by  visually
identifying about 20  2MASS stars in the field  under study. The stars
in the list  were then used to obtain  transformation equations to get
equatorial coordinates  for the remaining  stars. In a second  step, a
computer routine was used to  cross-identify all the sources in common
with the same catalogues by matching the equatorial coordinates to the
catalogued  ones.  The  rms of  the residuals  were  $\sim 0\farcs17$,
which is about the astrometric precision of the 2MASS catalogue ($\sim
0\farcs12$), as expected since  most of the coordinates were retrieved
from this catalogue.

\subsection{Final catalogue}

The above  procedure allowed us  to build an  astrometric, photometric
($UBVIJHK$) catalogue that constitutes the main observational database
used in  this study. A solid  analysis of the behavior  of the Stellar
Energy Distributions  (SEDs) can be  carried out with this  tool, thus
preventing  possible  degeneracies  in  the photometric  diagrams  and
allowing to  obtain more reliable  results. {\bf The full catalogue with
a total of 34310 stars is only available in electronic form at the Centre
de Donnes astronomiques de Strasbourg (CDS). It includes X--Y positions;
2MASS identification (when available); equatorial coordinates (epoch 2000.0);
optical and 2MASS photometry. Table~\ref{tab:data} is a summary of that
catalogue including only stars adopted as likely cluster members or probable
young stellar objects (YSOs).}

\begin{table*}
\caption{A sample of the final catalogue used in this study.}
\label{tab:data}
\fontsize{6} {8pt}\selectfont
\tabcolsep3pt
\begin{center}
\begin{tabular}{ccccccccccccccl}
\hline
ID & 2MASS~ID &
     \multicolumn{1}{c}{$X[pix]$}         & \multicolumn{1}{c}{$Y[pix]$}         &
     \multicolumn{1}{c}{$\alpha_{J2000}$} & \multicolumn{1}{c}{$\delta_{J2000}$} &
     \multicolumn{1}{c}{$V$}              & \multicolumn{1}{c}{$B$-$V$}          &
     \multicolumn{1}{c}{$U$-$B$}          & \multicolumn{1}{c}{$V$-$I$}          &
     \multicolumn{1}{c}{$K$}              & \multicolumn{1}{c}{$J$-$H$}          &
     \multicolumn{1}{c}{$H$-$K$}          & \multicolumn{2}{c}{Comments}       \\
\hline
%ID_new2     2MASS              x        y        RAJ2000       DEJ2000       v         bv        ub        vi        K         J-H       H-K       Zone
    256 &  J13122855-6241438  &  1662.1 &  2563.4 &   13:12:28.6 &  -62:41:43.4 &   14.964 &    2.448 &    1.389 &    3.720 &    6.610 &    0.989 &    0.656 & D1     & lm    \\
    659 &  J13122850-6241509  &  1660.7 &  2587.8 &   13:12:28.5 &  -62:41:50.5 &   16.214 &    2.417 &    1.426 &    3.854 &    8.150 &    0.979 &    0.684 & D1     & lm    \\
    733 &  J13122626-6242095  &  1607.8 &  2652.7 &   13:12:26.3 &  -62:42:09.2 &   16.359 &    2.466 &    1.283 &    3.679 &    8.507 &    0.906 &    0.499 & D1     & lm    \\
   1124 &  J13122617-6241576  &  1605.9 &  2611.5 &   13:12:26.2 &  -62:41:57.3 &   16.954 &    2.744 &    1.349 &    4.028 &    7.756 &    1.294 &    0.560 & D1     & lm    \\
   1527 &  J13122369-6242013  &  1546.9 &  2623.8 &   13:12:23.8 &  -62:42:00.8 &   17.355 &    2.771 &    1.772 &    4.519 &    7.531 &    1.218 &    0.665 & D1     & lm    \\
   1578 &  J13122497-6242002  &  1576.7 &  2619.6 &   13:12:25.0 &  -62:41:59.6 &   17.401 &    2.622 &    1.819 &    4.520 &    7.477 &    1.276 &    0.669 & D1     & lm    \\
   2683 &  J13122678-6241571  &  1621.4 &  2607.2 &   13:12:26.9 &  -62:41:56.0 &   18.160 &    2.406 &    1.905 &    4.019 &    8.135 &    2.076 &    0.566 & D1     & lm    \\
   2819 &  J13123412-6241359  &  1793.8 &  2535.6 &   13:12:34.1 &  -62:41:35.4 &   18.228 &    2.108 &    1.296 &    3.551 &   10.798 &    0.878 &    0.418 & D1     & lm    \\
   2861 &        --           &  1650.6 &  2551.2 &   13:12:28.1 &  -62:41:39.9 &   18.251 &    2.219 &    1.140 &    3.559 &      --  &      --  &      --  & D1     & lm    \\
   3335 &  J13122721-6242041  &  1630.7 &  2633.8 &   13:12:27.3 &  -62:42:03.7 &   18.480 &    2.157 &    1.604 &    3.795 &   10.046 &    1.058 &    0.637 & D1     & lm    \\
   3505 &        --           &  1601.0 &  2609.4 &   13:12:26.0 &  -62:41:56.6 &   18.562 &    2.091 &    1.942 &    3.891 &      --  &      --  &      --  & D1     & lm    \\
   3518 &        --           &  1670.5 &  2579.6 &   13:12:28.9 &  -62:41:48.1 &   18.568 &    2.184 &    1.542 &    3.691 &      --  &      --  &      --  & D1     & lm    \\
   3694 &  J13122288-6241488  &  1527.4 &  2580.8 &   13:12:22.9 &  -62:41:48.4 &   18.644 &    2.256 &    1.818 &    4.196 &    9.644 &    1.057 &    0.571 & D1     & lm    \\
   3921 &  J13122693-6242147  &  1624.0 &  2669.3 &   13:12:27.0 &  -62:42:14.0 &   18.737 &    2.129 &    1.515 &    3.811 &   10.296 &    0.856 &    0.685 & D1     & lm    \\
   4046 &        --           &  1608.4 &  2605.5 &   13:12:26.3 &  -62:41:55.5 &   18.780 &    2.380 &    1.763 &    4.019 &      --  &      --  &      --  & D1     & lm    \\
   4122 &  J13122705-6242079  &  1626.3 &  2646.4 &   13:12:27.1 &  -62:42:07.3 &   18.806 &    2.098 &    1.350 &    3.544 &    9.770 &    0.861 &    1.888 & D1     & lm    \\
   4168 &  J13122525-6241555  &  1585.2 &  2605.3 &   13:12:25.4 &  -62:41:55.5 &   18.826 &    2.155 &    1.894 &    3.837 &    8.904 &    2.692 &    0.338 & D1     & lm    \\
   4502 &  J13122959-6241291  &  1685.7 &  2512.4 &   13:12:29.6 &  -62:41:28.7 &   18.938 &    2.112 &    1.950 &    3.683 &   11.069 &    0.944 &    0.458 & D1     & lm    \\
   4664 &        --           &  1610.3 &  2639.3 &   13:12:26.4 &  -62:42:05.3 &   18.990 &    2.106 &    1.486 &    3.587 &      --  &      --  &      --  & D1     & lm    \\
   5245 &        --           &  1624.6 &  2633.6 &   13:12:27.0 &  -62:42:03.7 &   19.207 &    2.293 &    1.917 &    3.682 &      --  &      --  &      --  & D1     & lm    \\
   5557 &        --           &  1602.8 &  2621.1 &   13:12:26.1 &  -62:42:00.0 &   19.304 &    2.333 &    1.723 &    3.848 &      --  &      --  &      --  & D1     & lm    \\
   5569 &  J13123061-6242081  &  1710.3 &  2647.6 &   13:12:30.6 &  -62:42:07.7 &   19.309 &    2.326 &    2.661 &    4.129 &   10.222 &    1.118 &    0.616 & D1     & lm    \\
   5833 &  J13122454-6242088  &  1565.7 &  2649.8 &   13:12:24.5 &  -62:42:08.3 &   19.391 &    2.136 &    2.106 &    4.089 &   10.433 &    1.103 &    0.565 & D1     & lm    \\
   6519 &  J13122498-6241456  &  1577.3 &  2568.3 &   13:12:25.0 &  -62:41:44.8 &   19.589 &    2.278 &    1.565 &    3.799 &   11.286 &    0.950 &    0.594 & D1     & lm    \\
\hline
    223 &  J13125643-6240283  &  2325.1 &  2300.8 &   13:12:56.4 &  -62:40:27.8 &   14.793 &    2.386 &    1.617 &    3.547 &    6.873 &    0.896 &    0.527 & D2     & lm    \\
    264 &  J13124634-6241262  &  2085.4 &  2501.7 &   13:12:46.3 &  -62:41:25.7 &   14.999 &    2.324 &    1.272 &    3.406 &    7.792 &    0.818 &    0.425 & D2     & lm    \\
    287 &  J13124958-6241207  &  2162.7 &  2482.7 &   13:12:49.6 &  -62:41:20.3 &   15.095 &    2.120 &    1.051 &    3.127 &    8.595 &    0.686 &    0.400 & D2     & lm    \\
    775 &  J13125627-6240515  &  2321.6 &  2381.5 &   13:12:56.2 &  -62:40:51.1 &   16.409 &    2.343 &    1.105 &    3.489 &    9.306 &    0.770 &    0.402 & D2     & lm    \\
    892 &  J13125456-6241050  &  2280.9 &  2428.3 &   13:12:54.5 &  -62:41:04.6 &   16.618 &    2.343 &    1.057 &    3.388 &    9.654 &    0.779 &    0.369 & D2     & lm    \\
    936 &  J13125236-6240463  &  2228.5 &  2363.6 &   13:12:52.3 &  -62:40:45.9 &   16.672 &    2.159 &    1.070 &    3.205 &   10.059 &    0.725 &    0.369 & D2     & lm    \\
    979 &  J13125864-6240552  &  2378.1 &  2394.4 &   13:12:58.6 &  -62:40:54.9 &   16.728 &    2.349 &    1.059 &    3.491 &    9.556 &    0.803 &    0.396 & D2     & lm    \\
    989 &  J13125396-6240475  &  2266.4 &  2367.6 &   13:12:53.9 &  -62:40:47.1 &   16.741 &    2.196 &    1.056 &    3.209 &   10.104 &    0.731 &    0.334 & D2     & lm    \\
   1026 &  J13125445-6240459  &  2278.0 &  2362.3 &   13:12:54.4 &  -62:40:45.5 &   16.795 &    2.273 &    0.907 &    3.291 &    9.920 &    0.802 &    0.416 & D2     & lm    \\
   1260 &  J13125829-6240371  &  2369.8 &  2331.4 &   13:12:58.3 &  -62:40:36.7 &   17.098 &    2.340 &    1.154 &    3.611 &    9.628 &    0.820 &    0.413 & D2     & lm    \\
   1409 &  J13125325-6240346  &  2249.7 &  2323.0 &   13:12:53.2 &  -62:40:34.2 &   17.247 &    2.076 &    0.945 &    3.343 &   10.529 &    0.735 &    0.342 & D2     & lm    \\
   1577 &  J13124747-6240547  &  2112.2 &  2392.4 &   13:12:47.5 &  -62:40:54.2 &   17.399 &    2.163 &    0.898 &    3.271 &   10.725 &    0.756 &    0.326 & D2     & lm    \\
   1608 &  J13125436-6240422  &  2276.1 &  2349.2 &   13:12:54.3 &  -62:40:41.8 &   17.428 &    2.219 &    1.080 &    3.354 &   10.529 &    0.763 &    0.359 & D2     & lm    \\
   1631 &  J13125731-6240268  &  2346.2 &  2296.5 &   13:12:57.3 &  -62:40:26.6 &   17.453 &    2.327 &    1.048 &    3.582 &    8.504 &    1.789 &    0.836 & D2     & lm    \\
   1671 &  J13125373-6240508  &  2261.3 &  2379.1 &   13:12:53.7 &  -62:40:50.4 &   17.489 &    2.230 &    1.017 &    3.376 &   10.477 &    0.779 &    0.363 & D2     & lm    \\
   1704 &  J13125356-6240119  &  2257.1 &  2243.9 &   13:12:53.6 &  -62:40:11.4 &   17.513 &    2.355 &    1.254 &    3.725 &    9.891 &    0.854 &    0.413 & D2     & lm    \\
   1792 &  J13125529-6240416  &  2297.7 &  2346.7 &   13:12:55.3 &  -62:40:41.0 &   17.580 &    2.382 &    0.840 &    3.302 &   10.307 &    0.745 &    0.439 & D2     & lm    \\
   2248 &  J13130084-6239545  &  2430.5 &  2183.7 &   13:13:00.8 &  -62:39:54.0 &   17.918 &    2.134 &    1.356 &    3.287 &   11.229 &    0.777 &    0.308 & D2     & lm    \\
   2272 &  J13125269-6240545  &  2236.6 &  2391.7 &   13:12:52.7 &  -62:40:54.0 &   17.931 &    2.132 &    1.014 &    3.217 &   11.293 &    0.728 &    0.309 & D2     & lm    \\
   2500 &  J13125226-6240242  &  2226.0 &  2286.5 &   13:12:52.2 &  -62:40:23.6 &   18.073 &    2.115 &    1.056 &    3.379 &   11.066 &    0.766 &    0.394 & D2     & lm    \\
   2533 &  J13125193-6240579  &  2218.5 &  2403.7 &   13:12:51.9 &  -62:40:57.5 &   18.091 &    2.110 &    1.165 &    3.377 &   11.107 &    0.784 &    0.373 & D2     & lm    \\
   2700 &  J13125237-6240314  &  2229.1 &  2311.9 &   13:12:52.4 &  -62:40:31.0 &   18.168 &    2.071 &    1.086 &    3.279 &   11.344 &    0.741 &    0.404 & D2     & lm    \\
   3215 &  J13124526-6240421  &  2059.3 &  2349.1 &   13:12:45.3 &  -62:40:41.7 &   18.426 &    2.042 &    2.545 &    3.291 &   11.079 &    1.139 &    0.376 & D2     & lm    \\
   3388 &  J13125828-6240526  &  2369.5 &  2385.1 &   13:12:58.3 &  -62:40:52.2 &   18.506 &    2.031 &    1.354 &    3.370 &   11.505 &    0.793 &    0.340 & D2     & lm    \\
\hline
   3199 &  J13115353-6246577  &   829.8 &  3652.4 &   13:11:53.6 &  -62:46:57.5 &   18.420 &    1.874 &    1.475 &    2.775 &   11.702 &    0.976 &    0.704 & D130   & yso?  \\
   3486 &  J13115601-6247114  &   889.3 &  3700.3 &   13:11:56.1 &  -62:47:11.4 &   18.554 &    2.062 &    3.299 &    2.974 &   11.787 &    1.035 &    0.473 & D130   & yso?  \\
   3624 &  J13115851-6247072  &   948.6 &  3685.4 &   13:11:58.6 &  -62:47:07.1 &   18.617 &    1.962 &    2.159 &    3.004 &   11.700 &    1.087 &    0.407 & D130   & yso?  \\
   9376 &  J13115519-6247072  &   870.8 &  3685.7 &   13:11:55.4 &  -62:47:07.2 &   20.249 &    2.111 &    2.696 &    3.310 &   12.308 &    1.392 &   -0.259 & D130   & yso?  \\
  12300 &  J13115237-6246517  &   801.3 &  3633.2 &   13:11:52.5 &  -62:46:52.0 &   20.720 &    2.146 &    0.021 &    2.607 &   13.956 &    1.009 &    1.021 & D130   & yso?  \\
  13396 &        --           &   774.4 &  3636.3 &   13:11:51.3 &  -62:46:52.9 &   20.889 &    1.991 &    1.244 &    2.669 &      --  &      --  &      --  & D130   & yso?  \\
\hline
   7046 &        --           &   439.8 &   785.8 &   13:11:37.4 &  -62:33:09.7 &   19.737 &    2.531 &    3.769 &    4.343 &      --  &      --  &      --  & D131   & yso?    \\
   9892 &        --           &   479.9 &   800.9 &   13:11:39.1 &  -62:33:14.1 &   20.337 &    3.125 &    1.362 &    5.469 &      --  &      --  &      --  & D131   & yso?    \\
  12192 &        --           &   406.2 &   798.5 &   13:11:36.0 &  -62:33:13.4 &   20.704 &    2.326 &      --  &    4.493 &      --  &      --  &      --  & D131   & yso?    \\
  13636 &        --           &   450.4 &   800.1 &   13:11:37.9 &  -62:33:13.8 &   20.924 &    6.467 &      --  &    3.997 &      --  &      --  &      --  & D131   & yso?    \\
  14868 &  J13114106-6232570  &   525.7 &   741.5 &   13:11:41.1 &  -62:32:57.0 &   21.107 &    2.553 &   -0.300 &    5.166 &    9.575 &    1.366 &    0.819 & D131   & yso?-A1 \\
  15897 &        --           &   520.6 &   741.9 &   13:11:40.8 &  -62:32:57.1 &   21.247 &      --  &      --  &    5.334 &      --  &      --  &      --  & D131   & yso?    \\
  16128 &        --           &   622.8 &   805.9 &   13:11:45.1 &  -62:33:15.6 &   21.277 &    2.664 &    1.699 &    5.888 &      --  &      --  &      --  & D131   & yso?    \\
  16664 &        --           &   347.5 &   847.0 &   13:11:33.6 &  -62:33:27.3 &   21.351 &    2.534 &   -0.207 &    5.002 &      --  &      --  &      --  & D131   & yso?-A2 \\
  18863 &        --           &   534.4 &   812.4 &   13:11:41.4 &  -62:33:17.4 &   21.632 &    2.821 &    0.596 &    4.837 &      --  &      --  &      --  & D131   & yso?-CO \\
  21172 &  J13113947-6233283  &   487.9 &   850.0 &   13:11:39.5 &  -62:33:28.3 &   21.899 &    2.503 &    1.490 &    4.886 &   11.939 &    1.471 &    0.657 & D131   & yso?-A3 \\
  22042 &        --           &   417.2 &   791.1 &   13:11:36.5 &  -62:33:11.2 &   22.003 &      --  &      --  &    4.420 &      --  &      --  &      --  & D131   & yso?    \\
  25701 &        --           &   495.4 &   777.4 &   13:11:39.8 &  -62:33:07.3 &   22.452 &    2.685 &    0.942 &    4.791 &      --  &      --  &      --  & D131   & yso?    \\
  26479 &        --           &   496.5 &   818.1 &   13:11:39.8 &  -62:33:19.0 &   22.556 &      --  &      --  &    5.027 &      --  &      --  &      --  & D131   & yso?    \\
  28362 &        --           &   477.7 &   839.5 &   13:11:39.0 &  -62:33:25.2 &   22.873 &    2.629 &      --  &    5.341 &      --  &      --  &      --  & D131   & yso?    \\
  30644 &  J13113382-6233270  &   354.0 &   847.7 &   13:11:33.8 &  -62:33:27.5 &   23.475 &      --  &      --  &    3.694 &   10.342 &    1.308 &    0.678 & D131   & yso?    \\
\hline
\end{tabular}
\begin{minipage}{15cm}
Note 1: Comments column indicate the star region location and its adopted membership.
(D1/2 = Danks 1/2 regions; D130/131 = DBS130/131 regions; lm = likely member; yso? = probable YSO; A1, A2, A3 and CO are
identifications from Leistra et al. 2005) \\
Note 2: A full version of this table is only available in electronic form at the Centre de Donnes astronomiques de Strasbourg (CDS).
\end{minipage}
\end{center}
\end{table*}

\section{Data analysis} \label{sec:analysis}

\begin{figure*}
\centering
\centerline{\psfig{file=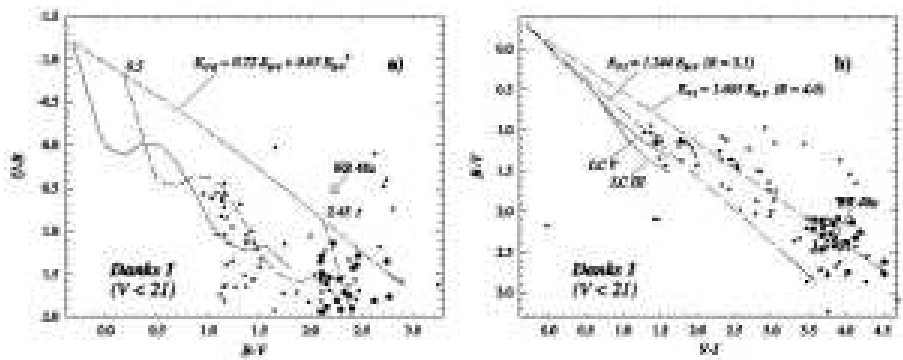,width=12cm}}
\caption{Optical  TCDs  of  stars  located  in $D1$  zone.   See  also
  Sect~\ref{sec:zones}  and Fig.~\ref{fig:dss}  for  zone definitions.
  {\bf a)}  $U-B$ vs.  $B-V$  diagram. Heavy black circles  are likely
  cluster members whereas light dots are likely field stars. The solid
  line  is the  Schmidt-Kaler (1982) ZAMS, while dashed lines are the
  same ZAMS, but shifted along the reddening line by the adopted color excesses
  indicated above them. They correspond to the adopted ones for the foreground
  population ($E_{B-V} = 0.5$) and for the cluster stars ($E_{B-V} = 2.45$).
  See also Sect.~\ref{sec:parameters}. The dashed arrow indicates the normal
  reddening path. {\bf b)} $B-V$  vs. $V-I$ diagrams. Symbols are the same  as
  in panel  a)  and solid  lines  are intrinsic colors for luminosity class V
  and  III from  Cousins (1978ab). Dashed arrows indicate the reddening paths
  for normal ($R_V = 3.1$) and abnormal $R_V$ values.}
\label{fig:d1-ccds}
\end{figure*}

\begin{figure*}
\centering
\centerline{\psfig{file=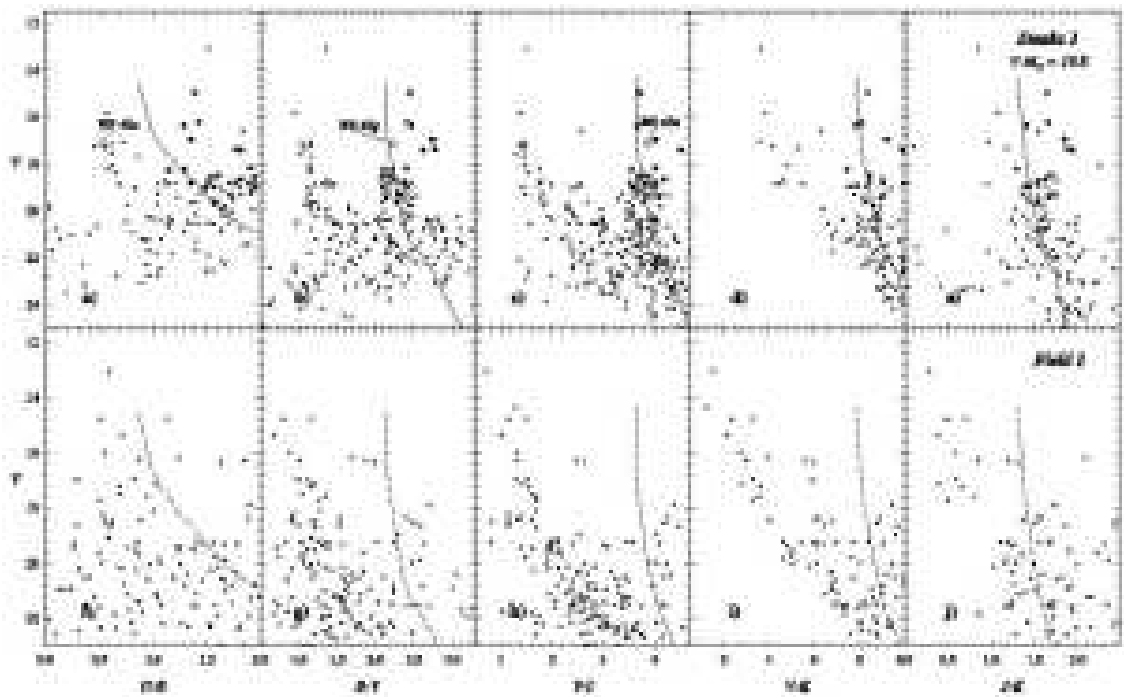,width=15cm}}
\caption{Optical CMDs  of stars  located in $D1$  and $F1$  zones. See
  also   Sect   ~\ref{sec:zones}   and  Fig~\ref{fig:dss}   for   zone
  definitions.    Symbols   are the    same    as    in
  Fig~\ref{fig:d1-ccds}.   The  solid  and   dashed  curves   are  the
  Schmidt-Kaler  (1982) empirical ZAMS  and the  MS path  from Cousins
  (1978ab)  and Koornneef (1983).  Solid curves  are corrected  by the
  adopted      cluster     apparent     distance      modulus     (see
  Sect.~\ref{sec:parameters}).}
\label{fig:d1-cmds}
\end{figure*}
\begin{figure*}
\centering
\centerline{\psfig{file=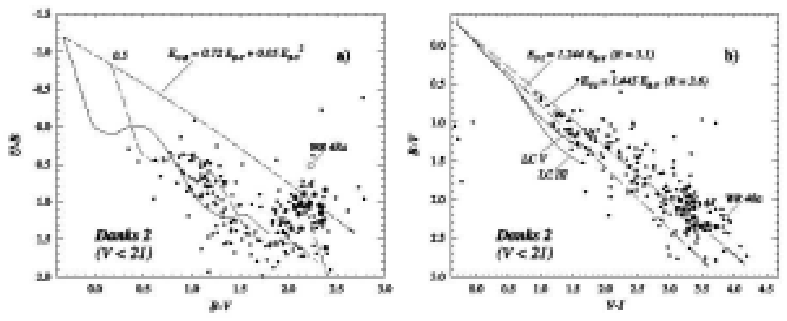,width=12cm}}
\caption{Optical TCDs of stars located in $D2$ zones. See also Sect~\ref{sec:zones}
and Fig.~\ref{fig:dss} for zone definitions. Symbols and curves are teh same
as in Fig.~\ref{fig:d1-ccds}}
\label{fig:d2-ccds}
\end{figure*}

\begin{figure*}
\centering
\centerline{\psfig{file=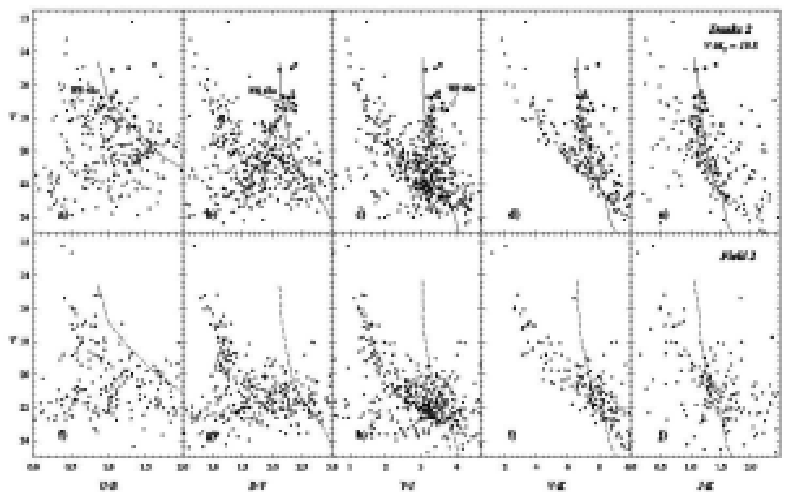,width=15cm}}
\caption{Optical CMDs of stars located in $D2$ and $F2$ zones. See also Sect~\ref{sec:zones}
and Fig~\ref{fig:dss} for zone definitions. Symbols and curves are the same
as in Fig~\ref{fig:d1-cmds}}
\label{fig:d2-cmds}
\end{figure*}

\subsection{Selection of zones} \label{sec:zones}

With  the aim  of  simplifying the  analysis  of the  data, the
studied   region  was   divided   into  several   sets  presented   in
Table~\ref{tab:zones} (see also Fig.~\ref{fig:dss}):

\begin{table}
\caption{{} Description of studied zones.}
%\fontsize{8} {10pt}\selectfont
\begin{center}
\begin{tabular}{lccc}
\hline
$Name$ & $Shape$ & $Center$     & $Radius~(r)$                \\
\hline
$D1$     & circle & Danks~1     & $r = 0\farcm96$             \\
$F1$     & corona & Danks~1     & $0\farcm96 < r < 1\farcm36$ \\
\hline
$D2$     & circle & Danks~2     & $r = 1\farcm44$             \\
$F2$     & corona & Danks~2     & $1\farcm44 < r < 2\farcm04$ \\
\hline
$DBS130$ & circle & DBS2003~130 & $r = 0\farcm72$             \\
$F130$   & corona & DBS2003~130 & $0\farcm72 < r < 1\farcm02$ \\
\hline
$DBS131$ & circle & DBS2003~131 & $r = 0\farcm72$             \\
$F131$   & corona & DBS2003~131 & $0\farcm72 < r < 1\farcm02$ \\
\hline
$Field$  & \multicolumn{3}{c}{The remaining observed area}    \\
\hline
\label{tab:zones}
\end{tabular}
\end{center}
\end{table}

In this  manner, $D1$, $D2$,  $DBS130$ and $DBS131$ correspond  to the
regions involving  the studied  clusters and their  surrounding areas;
while  $F1$, $F2$,  $F130$ and  $F131$ correspond  to  the respective,
adopted, comparison field  for each cluster. We emphasize two
facts: a) each comparison field covers the same sky area as the
corresponding cluster region, and b) as will be explained in
Sec.~\ref{sec:dbs}, embedded cluster BDS2003~132 was not included
in the present study.

\subsection{Photometric diagrams}

The Two Color  Diagrams (TCDs) and Color Magnitude  Diagrams (CMDs) of
the   different   zones    are   shown   in   Figs.~\ref{fig:d1-ccds}-
\ref{fig:field-cmds}   in  a   self  explanatory   format   (see  also
Sect.~\ref{sec:zones}).

The    TCDs   of    the   cluster    zones   (Figs.~\ref{fig:d1-ccds},
\ref{fig:d2-ccds} and \ref{fig:dbs-ccds}) show clearly the presence of
the,  expected,  field  population   along  with  a  heavily  reddened
($E_{B-V} \sim 2.5$)  group of stars, most likely  cluster members. In
all  studied  zones  $B-V$  vs.   $V-I$  diagrams  confirm  the  above
mentioned  trend and show  that the  clusters members  suffer abnormal
reddening laws, unlike the field population which follow a normal one.
As a consequencem individual $R$ ($=A_V/E_{B-V}$) values were
derived for each cluster.\\

The  optical  CMDs  of  the  cluster  zones  (Figs.~\ref{fig:d1-cmds},
\ref{fig:d2-cmds}  and \ref{fig:dbs-cmds})  also  illustrate that  the
cluster populations is  made of very reddened stars.  According to the
infrared CMDs,  2MASS data  have not enough  precision to  improve the
information obtained  from the optical  photometric diagrams. However,
it is  important to notice that  the chosen distance  for each cluster
(see Sect.~\ref{sec:parameters})  produced a coherent fit  of the ZAMS
or MS along all the CMDs  (optical and infrared) as lower envelopes of
the adopted member stars.

\subsection{Main objects in the studied region} \label{sec:parameters}

\begin{table*}
\caption{Parameters of the analyzed stellar groups}
\begin{center}
\begin{tabular}{lccccccccc}
\hline
Stellar group & \multicolumn{2}{c}{Center}        & $E_{B-V}$ & $R$       & $V-M_V$     & $A_V$     & $V_0 - M_V$   & Age[Myr])  & IMF slope  \\
              & $\alpha_{2000}$ & $\delta_{2000}$ &           &           &             &           &               &            &            \\
\hline
Danks~1       & 13:12:27.0      & -62:41:59.7     & 2.45      & 4.0       & 19.80       & ~9.8      & 10.0          & $\sim$ 5   & $\sim$ 1.5 \\
Danks~2       & 13:12:55.3      & -62:40:42.0     & 2.40      & 3.6       & 19.83       & ~8.7      & 11.5          & $\sim$ 5   & $\sim$ 1.1 \\
\hline
DBS2003~130   & 13:11:54.0      & -62:47:02.0     & 2.3       & $\sim$4.5 & $\sim$24.9  & 10.4      & 14.5          & $\sim$ 1-3 & -          \\
DBS2003~131   & 13:11:39.4      & -62:33:11.5     & 2.6       & $\sim$4.5 & $\sim$25.2  & 11.7      & 13.5          & $\sim$ 1-3 & $\sim$ 0.98 - 1.5 \\
\hline
Foreground    & 13:12:43.4      & -62:39:02.2     & $\sim$0.5 & 3.1       & $\sim$9-11  & $\sim$1.5 & $\sim$7.5-9.5 & -              & -      \\
\hline
\multicolumn{10}{l}{Note 1: Foreground coordinates correspond to the center of the observed area (see Fig.~\ref{fig:dss})} \\
\multicolumn{10}{l}{Note 2: IMF slope values for DBS2003~131 were taken from Leistra et al. (2005) and included for comparison.} \\
\label{tab:param}
\end{tabular}
\end{center}
\end{table*}

\subsubsection{Danks~1 and Danks~2}

Danks~1  (C1309-624) and  Danks~2 (C1310-624)  are catalogued  as open
clusters with  a diameter of $1\farcm0$  and $1\farcm5$, respectively,
and both  classified with a Trumpler class  1-1-p- (Lyng\aa~1987; Dias
et al.  2002).  That is why the sizes of the respective regions ($D1$,
$F1$,  $D2$  and  $F2$)  were chosen  (see  Sect.~\ref{sec:zones})  to
analyze the  cluster itself and  the behavior of its  associated field
respectively. Although  these clusters  are more easily  detectable in
infrared  still   they  appear   as  clear  over-densities   in  DSS-2
(Fig.~\ref{fig:dss}a). The adopted centers for the clusters were taken
from $SIMBAD$ and are presented in Table~\ref{tab:param}.  Photometric
diagrams  of these clusters (Figs.~\ref{fig:d1-ccds} - \ref{fig:d2-cmds})
show (i) they suffer substantial reddening ($E_{B-V} = 2.45$ for Danks~1
and 2.4 for Danks~2); (ii) exhibit abnormal $R$ value (3.6  -  4.0);
and  (iii)  may  also  suffer a differential reddening across selected
regions.
 In order to perform  membership assignment, the individual position
of the  stars  in  all  the  photometric  diagrams  have  been  carefully
inspected aiming at checking  their consistency in all these diagrams
simultaneously. This later point was performed assuming we were dealing
with young populations and the stars positions on the CMDs were close to
the location of a compatible ZAMS solution, by using previously adopted
color excess ratios and color excess values. However, we allow for some
dispersion around the adopted ZAMS due to the probable presence of dust
inside the cluster, and also due to photometric errors (mainly in $U$ band).
As a following step, we took into account the number of stars for each
magnitude bin according to the apparent LFs (see Baume et al. 2004ab, 2006).
This procedure was applied for all stars down  to an adopted $V_{lim}\sim 19$.
At fainter magnitudes, contamination by field stars becomes severe, preventing
an easy identification of faint cluster members. The fit of a properly reddened
Schmidt-Kaler (1982) ZAMS to the blue edge of the adopted member stars yields a
distance modulus $V_0 - M_V = 10.0 \pm 0.3$ for Danks~1 and $V_0 - M_V = 11.5 \pm 0.3$
for Danks~2 (errors from eye inspection). An age of about 5 Myr can be estimated
for both clusters taking into account Meynet et  al. (1993) calibration. It must
be noticed that distance  values are lower than those obtained in previous works
(e.g.  Bica et al.  2004) and this is mainly due to the abnormal extinction
law that we derive and adopt in the present study.

To  compute the  clusters LFs  and  IMFs the  following procedure  was
adopted:

\begin{itemize}
\item Stars located in cluster zones ($D1$ and $D2$) and comparison
  zones ($F1$ and $F2$) were selected. Additionally, in order minimize the
  contamination of stars from the field population, only those stars with
  $V-I > (V-I)_{lim}$ were considered (values of $(V-I)_{lim}$ equal to 3.4
  and 3.0 were adopted for $D1/F1$ and $D2/F2$ respectively). All these stars
  were called then "red stars"
\item The apparent LFs of the clusters were obtained substracting the
  apparent LF of "red stars"  in the corresponding comparison zone ($F1$
  or $F2$)  from the apparent LF  of "red stars" placed  in the clusters
  zones ($D1$ or  $D2$).  Bins were chosen in a  way to avoid negative
  final values.
\item The resulting apparent LFs  were shifted in magnitude to correct
  them by absorption and distance and to obtain the final LFs.
\item Finally, the  IMFs of the clusters were computed converting LFs
  bins  into mass  bins using  the mass-luminosity  relation  given by
  Scalo (1986).
\end{itemize}

\begin{figure}
\centering
\centerline{\psfig{file=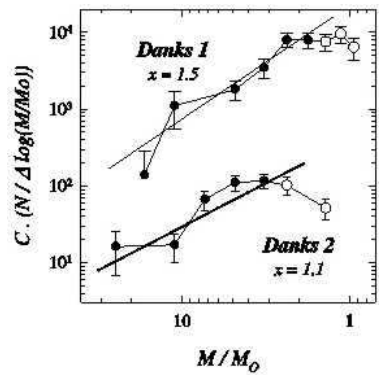,width=5cm}}
\caption{Initial  Mass Functions  (IMFs) of  Danks~1 (upper  plot) and
  Danks~2 (lower  plot). Error bars  are from Poisson  statistics. The
  least square  fittings for  the more massive  bins are  indicated by
  solid  right lines  (open  circles  indicate bins  not  used in  the
  fits. See text for details). For clarity, each IMF was shifted by an
  arbitrary constant (``$C$'').}
\label{fig:imfs}
\end{figure}

The   resulting    apparent   LFs   and   IMFs    are   presented   in
Table~\ref{tab:alf}  and  Fig.~\ref{fig:imfs}  respectively,  together
with the corresponding IMFs slopes.

\begin{figure*}
\centering
\centerline{\psfig{file=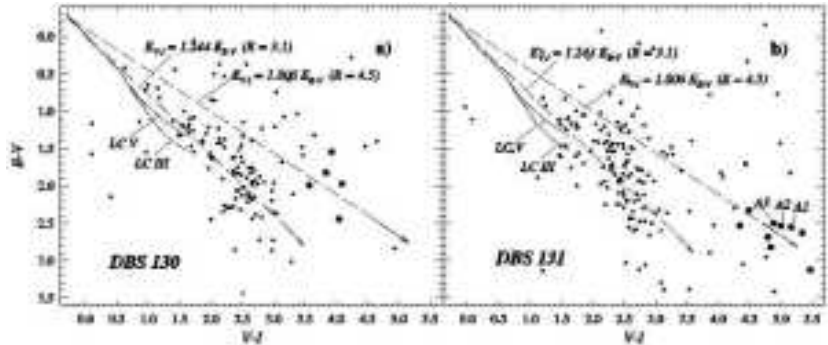,width=11cm}}
\caption{Optical  TCDs  of  stars  located in  $DBS130$  and  $DBS131$
  zones. See also Sect.~\ref{sec:zones} and Fig~\ref{fig:dss} for zone
  definitions.  Symbols   and  curves are the same   as  in
  Fig.~\ref{fig:d1-ccds}}
\label{fig:dbs-ccds}
\end{figure*}

\begin{figure*}
\centering
\centerline{\psfig{file=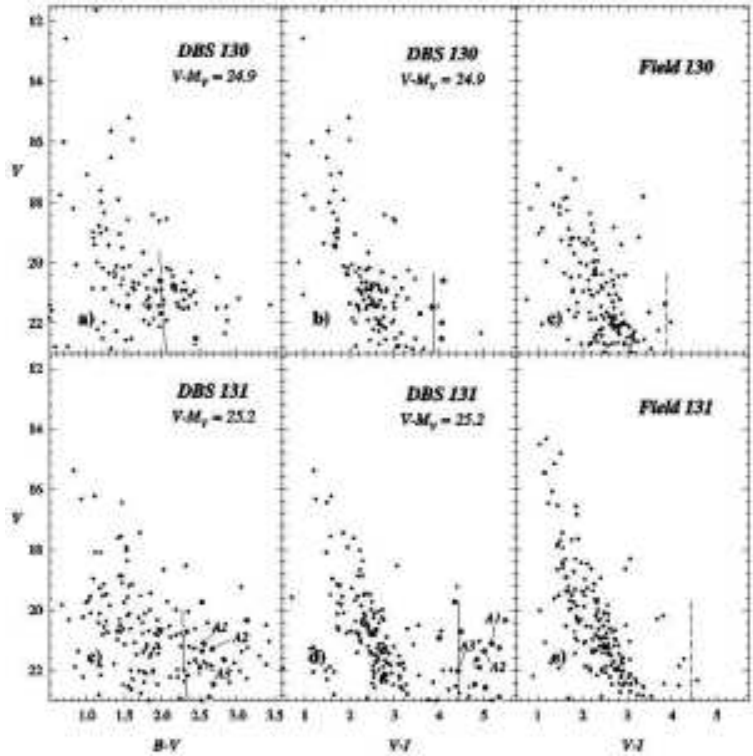,width=10cm}}
\caption{Optical CMDs  of stars located in  $DBS130$, $F130$, $DBS131$
  and    $F131$   zones.     See   also    Sect.~\ref{sec:zones}   and
  Fig~\ref{fig:dss} for zone definitions.   Symbols and curves are the
  same  as in  Fig~\ref{fig:d1-cmds}. $A1$,  $A2$ and  $A3$  are stars
  identification from Leistra et al. (2005).}
\label{fig:dbs-cmds}
\end{figure*}

\begin{table}
\caption{{} Apparent Luminosity Functions.}
%\fontsize{8} {10pt}\selectfont
\begin{center}
\begin{tabular}{cccc}
\hline
\multicolumn{2}{c}{Danks~1} & \multicolumn{2}{c}{ Danks~2} \\
$\Delta V$ & {\it N}        & $\Delta V$ & {\it N}         \\
\hline
14.0-16.0  & ~1             & 14.0-16.0  & ~3 \\
16.0-18.0  & ~4             & 16.0-17.0  & ~6 \\
18.0-19.0  & 13             & 17.0-18.0  & 12 \\
19.0-20.0  & 12             & 18.0-19.0  & 20 \\
20.0-21.0  & 22             & 19.0-20.0  & 20 \\
21.0-22.0  & 20             & 20.0-22.0  & 14 \\
22.0-23.0  & 16             & 22.0-24.0  & 12 \\
23.0-24.0  & 17             & 24.0-26.0  & ~0 \\
24.0-25.0  & 10             &            &    \\
\hline
\label{tab:alf}
\end{tabular}
\end{center}
\end{table}

\subsubsection{Embedded clusters} \label{sec:dbs}

The  region under  study contains  three recently  detected  (Dutra et
al. 2003) embedded clusters.  They were identified as BDS2003~130, 131
and  132. Each  cluster is  respectively found  very near  to  the HII
regions  G305.27-0.01, G305.3+0.2  and G305.3+0.1,  which can  also be
identified respectively with the S155,  S156 and S154 bubbles from the
Churchwell et al. (2006) catalogue.

Our  observations  allowed  us  to  obtain  reliable  information  for
BDS2003~130  and 131.  Unfortunately  BDS2003~132 is  too near  to the
position of star HD~114515  preventing to reach faint magnitudes. Both
BDS2003~130  and  BDS2003~131  (=   G305.3+0.2)  are  not  visible  in
Fig.~\ref{fig:dss}a, and their presence  is suggested only in our deep
frames  (see Fig.~\ref{fig:dss}b  and  d).
Nevertheless, the CMDs of  their zones ($DBS130$ and $DBS131$) clearly
trace their  presence, especially when the diagrams  are compared with
their  respective comparison fields  ($F130$ and  $F131$) as  shown in
Fig~\ref{fig:dbs-cmds}.         Additionally,        their        TCDs
(Fig.~\ref{fig:dbs-ccds})  reveals   that  both  cluster   suffers  an
abnormal  extinction law  and  a $R  \sim  4.5$ value  is adopted  for
them. It must be noticed  that since almost all the considered cluster
stars are at the limit of our photometry, their individual values must
be taken  as preliminary estimations  and only the group  structure in
the diagrams  can be considered  valid.  Still, stars  indicated with
black circles  in the  photometric diagrams can  be considered as very
probable YSOs.
In  this way, it is possible to obtain an estimation of their parameters
considering they are suffering a similar reddening than Danks~1/2
($E_{B-V} \sim  2.5$).  This assumption is coherent  with the $B-V$
values ($\sim  2.1-2.4$) of assumed  YSOs members.  To  estimate their
distances,  we again  use a  shifted Schmidt-Kaler  (1982)  ZAMS.  The
obtained clusters  parameters are presented  in Table~\ref{tab:param}.
In  the case  of BDS2003~131,  the  distance is  comparable with  that
derived in other studies (Leistra et al.  2005; Longmore et al.  2007)
based on near  infrared data or kinematic modeling  of the related HII
region (see Churchwell et al. 2006).

\begin{table}
\caption{{} Photometric data and adopted parameters for WR 48a.}
\begin{center}
\begin{tabular}{lc@{~~~~~~}r@{$~\sim~$}l}
\hline
\multicolumn{2}{c}{Photomety} & \multicolumn{2}{c}{Parameters} \\
\hline
$U$ & 19.89                   & $E_{B-V}$     & 2.4  \\
$B$ & 19.37                   & $R_{ISM}$     & 3.8  \\
$V$ & 17.13                   & $(A_V)_{ISM}$ & 9.1  \\
$I$ & 13.25                   & $(A_V)_{CSM}$ & 2    \\
$J$ & ~8.74                   & $V_O-M_V$     & 10.9 \\
$H$ & ~6.80                   & $M_V$         & -4.9 \\
$K$ & ~5.09                   & \multicolumn{2}{c}{} \\
\hline
\label{tab:wr}
\end{tabular}
\begin{minipage}{5cm}
\fontsize{7} {9pt}\selectfont
Note 1: $JHK$ photometry was taken from 2MASS catalogue \\
Note 2: $M_V$ was taken from Lundstr\"om \& Stenholm (1984)
\end{minipage}
\end{center}
\end{table}

\subsubsection{Star WR 48a}

WR 48a is a WC9 star (Danks et al. 1983) and it is placed within $\sim
1'$ of the  clusters Danks 1/2.  The distance  estimation performed by
Danks et al. (1983) was of 4 kpc, however van der Hucht (2001) concluded
it was only about 1.2 kpc away. On the other hand, Danks et al. (1983)
estimation of the absorption that this star is suffering is $A_V  \sim 9.2$.
This value is compatible with the values found for Danks~1/2 (see
Table~\ref{tab:param}).
Notwithstanding WR stars can involve circumstellar envelopes producing
intrinsic absorption, seems reasonable to associate WR 48a star to the
clusters  Danks~1/2 distance.  Given  the photometric  values obtained
for this star (see Table~\ref{tab:wr}) and its position in the CMDs of
the  clusters  it results  compatible  with  being  probable run  away
member as was previously assumed by Lundstrom \& Stenholm (1984) In this
case the  average values  of the  parameters between Danks~1/2 can be
assumed for the  WR star. In this way, the absorption value  comes only
from  the interstellar  medium ($(A_V)_{ISM}$) and adopting also
$M_V  = -4.9$ value from Lundstr\"om  \& Stenholm (1984) for a WC9 star,
it is possible to estimate the circumstellar absorption ($(A_V)_{CSM}$).
All the obtained values are presented in Table~\ref{tab:wr} and they
indicate that only a minor amount of the total visual absorption is
produced by the stellar envelope.

\subsection{The Field}

\begin{figure*}[t]
\centering
\centerline{\psfig{file=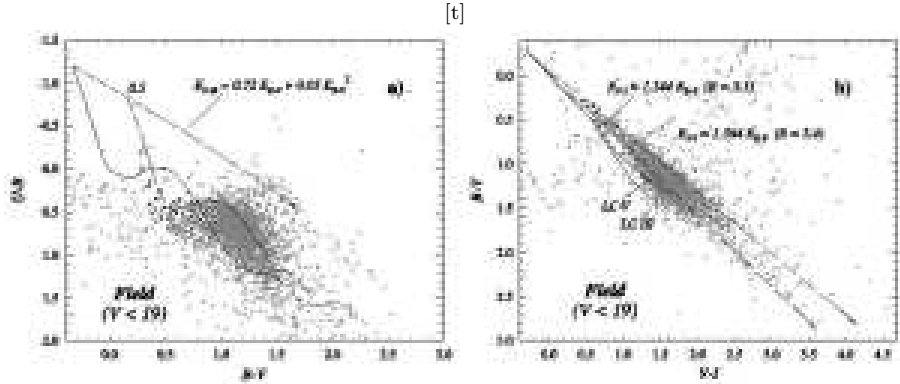,width=12cm}}
\caption{Optical  TCDs of  stars  located in  $Field$  zone. See  also
  Sect~\ref{sec:zones}  and  Fig~\ref{fig:dss}  for zone  definitions.
  Curves are the same as in Fig~\ref{fig:d1-ccds}.}
\label{fig:field-ccds}
\end{figure*}

\begin{figure*}
\centering
\begin{tabular}{cc}
\psfig{file=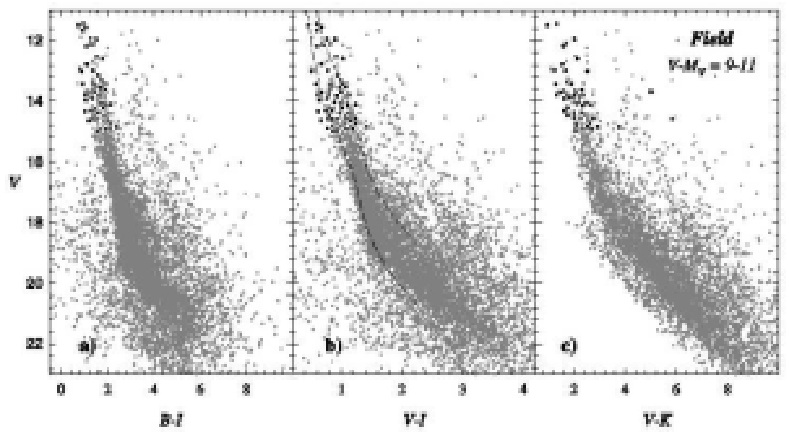,width=10cm,height=6cm} &
\psfig{file=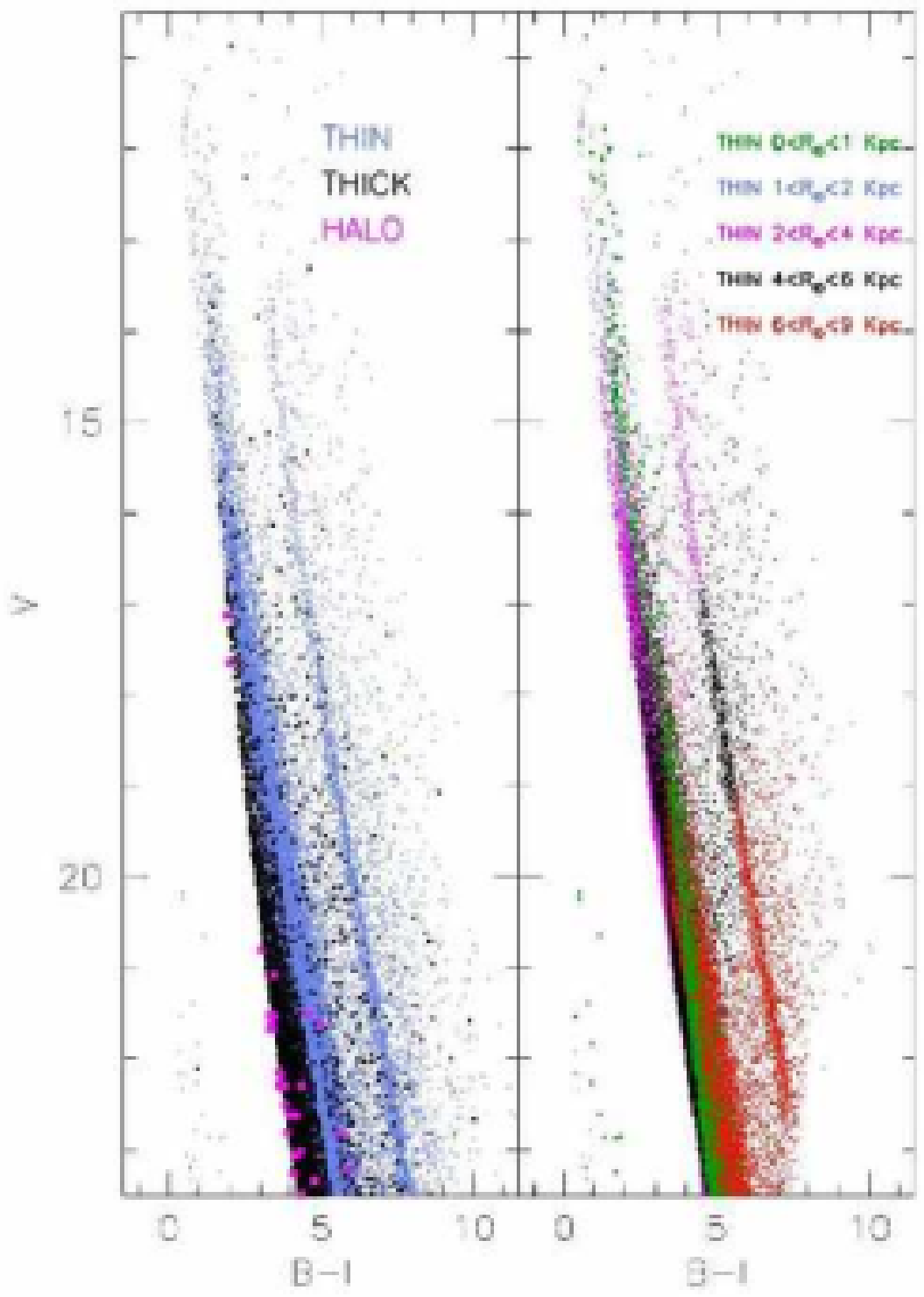,width=7cm,height=6cm} \\
\end{tabular}
\caption{Left plots: Observational optical CMDs of stars located in
  $Field$ zone, see Sect.~\ref{sec:zones} and Fig~\ref{fig:dss} for zone
  definitions. Curves are the  same as in Fig.~\ref{fig:d1-cmds}. Right
  plots: A simulation of the CMDs corresponding to the Galactic contribution
  along the ($l,b$)$=$($305,37^{\circ} +0.12^{\circ}$) line of
  sight. The left panel highlights the different contribution of the
  thin/thick disk and halo populations and the right panel shows the
  same  simulation but  analyzing the  distribution of  the  thin disk
  population at different distances along  the line of sight. The red
  oblique  sequence is  best explained  as  due to  the projection  of
  giants and red clump stars at different distances.}
\label{fig:field-cmds}
\end{figure*}

The  comparison of  the photometric  diagrams (TCDs  and CMDs)  of the
field   region   are   presented   in   Figs.~\ref{fig:field-ccds}   -
\ref{fig:field-cmds}. The $U-B$ vs. $B-V$ diagram reveals the presence
of a  relatively small  group of blue  stars (black symbols).   On the
other  hand,  all  the CMDs  show  two  clear  parallel blue  and  red
sequences.

For a deeper  understanding of these diagrams, and  in particular that
of  the  parallel red  sequence,  we  make  a comparison  between  the
observed diagrams and that of the expected Galactic contribution along
this line of sight.  The later is estimated by making use of the Robin
et al.   (2003) Galactic  model which provides  synthetic CMDs  at any
given  Galactic  coordinates.   These  synthetic  diagrams  have  been
successfully  used (e.g.  Momany et  al.  2004, 2006)  to explain  the
presence of  other seemingly  anomalous features around  Galactic open
clusters.
Figure~\ref{fig:field-cmds}  (see the  colored  version) compares  the
observed  $V$ vs. $B-I$  CMD with  the expected  Galactic contribution
along the ($l$,$b$) = ($305.37^{\circ}$,$+0.12^{\circ}$) line of sight
according  to the  {Besan\c{c}on} simulation  \footnote{Note  that the
  synthetic diagrams do not suffer completeness effects}.
In the  middle panel of the right  plot (of Fig.~\ref{fig:field-cmds})
we first disentangle between  the stellar populations belonging to the
three simulated  Galactic main components  (thin and thick  disks plus
halo) in this direction.  Clearly, the simulation shows that we expect
little,  if any,  halo contribution,  and  that most  of the  Galactic
contribution is due to thin  disk populations.  The true nature of the
red and oblique  sequence is best explained in the  right panel of the
plot.  Indeed,  when analyzing the  distance distribution of  the thin
disk  populations,  the  location  of  the  red  oblique  sequence  is
re-constructed as the projection of  red clump stars and bright giants
at different distances.

Additionally, the use of  the Schmidt-Kaler ZAMS (1982) indicates that
the brighter part of the  blue sequence (black symbols in left panel
of  Fig~\ref{fig:field-cmds}) is  associated  with a  nearby group  of
stars  with $E_{B-V}  \sim 0.5$  and  a distance  modulus $V-M_V  \sim
9-11$, corresponding to about 350-750 pc.

\section{Discussion} \label{sec:discussion}

\subsection{The Galactic spiral structure in Carina and Centaurus regions}

Previous investigations  by our group (V\'azquez et  al. 2005; Carraro
\& Costa 2009) presented optical observations near the field addressed
in  this  paper.   The  multiple young and blue  populations
along the line of view was found  to
reveal reveal three different spiral features at increasing distance from the Sun..

In  this paper, we strengthen this  conclusion  and lend further support
to the presence of at least three populations in this Galactic sector:
(i) the  first is a  spatially
spread foreground  population placed  mainly between 350-750  pc; (ii)
the second is situated just behind a very dark cloud at  1-3 kpc and is
represented by the clusters Danks~1/2; (iii) and, finally, a third population
which is traced by the embedded clusters, located at about 5-7 kpc.

\begin{figure*}
\centering
\centerline{\psfig{file=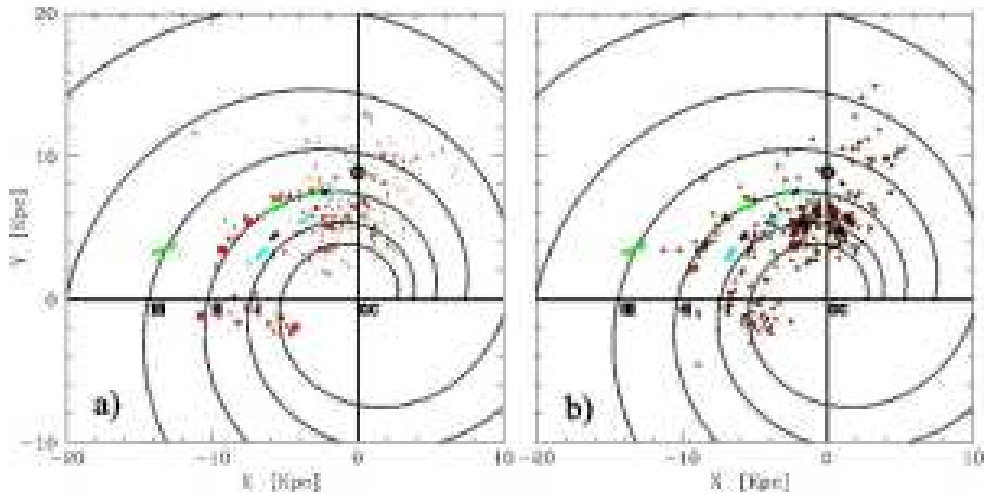,width=13cm}}
\caption{ The groups we discuss in this paper are here superposed to
  the Vall\'ee (2005) model of Galactic spiral structure (dotted curves)
  together with some of the HII regions (red symbols) adapted from Russeil (2003)
  in left panel (a) and Paladini et al. (2004) in right panel (b). Our groups
  are indicated as solid black segments representing their location and distance
  error. The green groups have been discussed in  Carraro \& Costa (2009) and are
  located at $l=290^\circ$, whilst the blue groups have been studied by V\'azquez
  et al. (2005) toward $l = 306^\circ$. The symbols I, II and III indicate the
  Scutum-Crux, Carina-Sagittarius and Perseus arms, respectively. See text for
  more details.}
\label{fig:arms}
\end{figure*}

The results are illustrated  in Fig.~\ref{fig:arms} where the Vall\'ee
(2005) spiral arms' model is also shown as dotted curves. In this  plot we
summarized the recent findings of Carraro \&  Costa (2009), V\'azquez et al.
(2005) and  the present study. The various the segments indicate the
location and corresponding uncertainties of the  different populations detected
along the $l = 290^{\circ}$, $305^{\circ}$ and $306^{\circ}$ line of  sight.

We confirmed the early findings by V\'azquez  et al.  (2005)  that the
populations beyond the open cluster Stock~16 (their groups {\bf A} and
{\bf  B})  are  most  probably  associated to  Scutum-Crux.  The  same
association could tentatively be done for the embedded clusters studied
in  this paper.  The Danks~1/2  clusters, on  the other  hand, clearly
belong  to  the Carina  branch  of  the  Carina-Sagittarius arm,  like
Stock~16  and the  population {\bf  A} described  in Carraro \& Costa
(2009). In this scenario it appears however that the Scutum-Crux arm
as  traced  by  Vall\'ee should have a larger pitch angle than 11$^{\circ}$.
In fact his plots indicate this parameter has a $\sigma \approx 4^{\circ}$.
However, some alternative scenarios are viable: (i) the observed displacement
from the model can be accounted for by the natural width of 1 Kpc for spiral arms
(see Vall\'ee 2005), or (ii) the  two  most distant groups falling in
between Scutum-Crux  and Carina  are tracing  an inter-arm structure,
or a branch of Carina. This later possibility finds some support from
the existence of several HII regions at the same position of these two
groups (Russeil 2003; Paladini 2004; see also Fig.~\ref{fig:arms}).

More directions close in this regions have to be studied. For example,
more deeper, detailed and complementary analysis in different wavelengths
as optical and infrared by using objects as embedded clusters seems necessary
to better trace the spiral structure in the fourth Galactic quadrant.

\subsection{Cluster IMFs}

There is an  important spread in the IMFs  slopes computed for massive
stars among several young cluster in the Galaxy. This is not yet fully
understood, and can be attributed to the fact that (a) the IMF may not
have a universal shape or/and  (b) there are intrinsic mistakes in its
computation (Scalo 1998).

Considering then the IMF slope given by:\\

$x = -log(dN/dlogM)/logM$,\\

\noindent
where $dN$ is the number of stars within the logaritmic mass
interval $dlogM$ around $logM$. The widely accepted slope values are
$x = 1.35$ (Salpeter 1955) or $x = 1.7$ (Scalo 1998). However, in several
very young open clusters (age $< 10^7$ yr, see e.g. Baume et al.
1999, 2003, 2004ab and references therein) we detect lower values
for this parameter.

Our computed values for the clusters Danks~1/2 together with the value
obtained by Leistra et al. (2005) for DBS2003~131 are presented in
Table~\ref{tab:param}. The results are coherent and lie within the
range of typically accepted values for very young clusters. However, we remark
that our derived IMFs and their slopes values are influenced by poor
statistics.

\section{Conclusions} \label{sec:conclusions}

A uniform  and deep study  was performed in  a region of  the Galactic
plane  towards  $l =  305^{\circ}$.   This  region  includes the  open
clusters Danks~1/2, their surrounding field and also three embedded IR
clusters.  The  main properties  of the different  populations located
along  this  line   of  sight  were  derived  and   a  description  of
interstellar medium (through the color excess behavior) has allowed us
to sketch a  better picture of the observed  clusters in this Galactic
direction.
In  particular, the  basic  parameters of  Danks~1/2, DBS2003~130  and
DBS2003~131 were determined.  It must  be noticed that the method used
analyzing star by star and  the inclusion of near ultraviolet $U$ data
has helped us  obtain mean distance values for  the open clusters that
are quite  different from  those derived in  recent papers,  where the
latter are based on the analysis  of only the global morphology of the
CMDs. Additionally, our optical methods applied for classical embedded
clusters offer compatible results with those especific ones used in
infrared, giving then more reliability to the results achieved in both
cases.
Our results indicate that, in order to obtain  better picture of complex
and reddened  regions, near ultraviolet data  are necessary. The performed
analysis indicates that Danks~1/2 are almost at the same distance from
the Sun, located probably on the Scutum-Crux arm. The analysis of the
location  of these grouping, together  with previous detections in the
same portion of the disk, are helping us  to get a better picture of the
spiral structure of the inner Milky Way.

Additionally, rough  estimates of  the  IMFs  for  both objects  were
computed for the first time. The derived slope values were similar to
those of other studied addressing young open clusters. Lastly, the
optical  photometric values of WR 48a were presented and its main
parameters were estimated, considering it as a runaway member of
Danks~1/2.

\section*{Acknowledgments}
GB acknowledges support from CONICET (PIP~5970) and the staff of CTIO during all the run of the observations
performed in March 2006. The authors are much obliged for the use of the NASA Astrophysics Data System, of the
$SIMBAD$ database (Centre de Donn\'es Stellaires --- Strasbourg, France) and of the WEBDA open cluster database.
This publication also made use of data from the Two Micron All Sky Survey, which is a joint project of the
University of Massachusetts and the Infrared Processing and Analysis Center/California Institute of Technology,
funded by the National Aeronautics and Space Administration and the National Science Foundation.

\end{document}